\documentclass[aip,jmp,amsmath,amssymb,onecolumn]{revtex4-1}
\newcommand{\so}{\text{\tiny(2)}}
\begin{document}
\title{Level density of a Bose gas: beyond the saddle point approximation}

\author{A.~D.~Ribeiro}
\affiliation{Departamento de F\'{\i}sica, 
Universidade Federal do Paran\'a, C.P. 19044, 
81531-980, Curitiba, PR, Brazil}

\date{\today}

\begin{abstract}

The present article is concerned with the use of approximations in the calculation of the many-body density of levels $\rho_{\mathrm{mb}}(E,N)$ of a system with total energy $E$, composed by $N$ bosons. In the mean-field framework, an integral expression for $\rho_{\mathrm{mb}}$, which is proper to be performed by asymptotic expansions, can be derived. However, the standard second order steepest descent method cannot be applied to this integral when the ground-state is sufficiently populated. Alternatively, we derive a uniform formula for $\rho_{\mathrm{mb}}$, which is potentially able to deal with this regime. In the case of the one-dimensional harmonic oscillator, using results found in the number theory literature, we show that the uniform formula improves the standard expression achieved by means of the second order method.

\end{abstract}


\maketitle

\section{Introduction}
\label{intro}

The many-body density of states $\rho_{\mathrm{mb}}(E,N)$ is the function that represents the number of possibilities of sharing the total energy $E$ of a physical system among its $N$ particles. Its evaluation for a general Hamiltonian is not an easy task, but it becomes considerably simpler in the mean-field approximation. The reason for such a simplicity is because the computation of the many-body density of states, in this case, reduces to the problem of finding the number of ways into which the energy $E$ can be shared among the $N$ particles occupying the individual levels of the mean-field potential. Then, the single-particle spectrum $\rho_{\mathrm{sp}}(\epsilon)$ of this potential becomes the basic ingredient to evaluate $\rho_{\mathrm{mb}}$.

Considering this framework, by using standard methods of statistical mechanics, a link between $\rho_{\mathrm{mb}}$ and $\rho_{\mathrm{sp}}$ can be easily delineated and an integral expression for the many-body density of states is straightforwardly achieved. Assuming, additionally, that values for $E$ and $N$ are sufficiently large, which points out to a connection with the thermodynamical limit, the integral representation of $\rho_{\mathrm{mb}}$ becomes proper to be performed by means of the saddle point method (or steepest descent method)~\cite{bleistein}. In this scenario, if one considers that there is only one saddle point isolated from other critical points, like extremes of the path of integration or singularities, the calculation simply consists in evaluating contributions of its neighborhood. In general, a second order expansion of the integrand around the saddle point is enough to accomplish this task, so that a final formula $\rho_{\mathrm{mb}}^{\so}$ for the many-body density of states is easily reached. Essentially, such a formalism has been satisfactorily used in the description of both bosonic~\cite{mbdsB2} and fermionic~\cite{mbdsF1,mbdsF2} systems.
 
A critical discussion concerning the evaluation of $\rho_{\mathrm{mb}}^{\so}$ leads us to conclude that this approach fails when the saddle point goes to zero. It can be asserted because, in this case, the saddle point approaches to a singularity of the integrand. When it happens, however, the method used to perform the integral can still be improved in order to avoid undesired effects. Actually, a uniform approximation~\cite{bleistein} can be derived, which correctly contemplates this particular situation. Roughly speaking, the new method consists in establishing a convenient mapping between the original (and complicated) integrand and a simpler one, where the integration process is workable. Our goal here is to study this scenario, deriving a uniform approximation $\rho_{\mathrm{mb}}^{\mathrm{un}}$ for the many-body density of states.

In order to check the accuracy of the uniform formula, we apply the formalism to non-interacting bosons confined to the one-dimensional harmonic oscillator (1D-HO). The reason for that is the close relation between $\rho_{\mathrm{mb}}(E,N)$, for this case, and the well-known problem of integer partition. As the single-particle levels are equidistant, by taking $E$ also as an integer, such a physical quantity becomes precisely the number of ways into which $E$ can be expressed as a sum of, at most, $N$ partitions. In number theory literature, it was solved by Erdos and Lehner in 1941~\cite{erdos}, in the asymptotic limit of large values of $E$ and $N$. Also, for an unrestricted number of partitions, which can be achieved by taking $N\to\infty$ and fixed $E$, this problem had already been solved by Hardy and Ramanujan~\cite{hardy} about one hundred years ago. 

As we show here, the uniform approximation satisfactorily improves the equivalent result given by $\rho_{\mathrm{mb}}^{\so}$, for the 1D-HO case, large $N$ and $E$, but $E\gg N$. However, there is an important discrepancy present in the pre-exponential term of $\rho_{\mathrm{mb}}^{\mathrm{un}}$: the $E$-dependence is a power law, as well as the Erdos and Lehner formula, but their exponents do not match. We then demonstrate that this difference vanishes with an additional consideration. Actually, the first result is achieved by providing the single-particle spectrum by means of the Thomas-Fermi method~\cite{brack}, which basically consists in evaluating the classically allowed phase-space region for a given energy $\epsilon$, associating it to $\rho_{\mathrm{sp}}(\epsilon)$. If we replace this method by a proper use of the Dedekind function~\cite{dedekind} and its properties, we recover the Erdos and Lehner formula, except for minor differences involving numerical factors.

This paper is organized as follows. In Sect.~\ref{formalism}, we introduce the formalism in which $\rho_{\mathrm{mb}}$ is linked to $\rho_{\mathrm{sp}}$ by means of an integral representation for $\rho_{\mathrm{mb}}$. In Sect.~\ref{spm}, we briefly evaluate the many-body density of states using the second order saddle point method, while, in Sect.~\ref{un}, the uniform formula $\rho_{\mathrm{mb}}^{\rm un}$ is derived. In Sect.~\ref{smo}, we apply our formalism to cases where $\rho_{\mathrm{sp}}$ is assumed to be given by the Thomas-Fermi method in order to perform a comparison (Sect.~\ref{OHS}) with other results found in the literature. Finally, in Sect.~\ref{fr}, we present our concluding remarks.

\section{Integral expression for $\rho_{\mathrm{mb}}$}
\label{formalism}

In this section we present the formalism that we will use to introduce approximations in the level density of Bose gases. It simply consists in finding an integral representation for $\rho_{\mathrm{mb}}$, depending just on $E$, $N$, and the single particle spectrum $\rho_{\mathrm{sp}}$, where the only assumption is that particles interact via a mean-field potential. The integral expression enables the use of the saddle point method~\cite{bleistein} to solve it.

The density of states $\rho_{\mathrm{mb}}(E,N)$ of a system composed of $N$ particles with total energy $E$ can be written as
\begin{equation}
\begin{array}{lll}
\rho_{\mathrm{mb}}(E,N) &=& \displaystyle 
\sum_{\nu}\delta(N-N_\nu)~\delta(E-E_\nu) 
\\ &=& \displaystyle 
\sum_{\nu} e^{u(N-N_\nu)}e^{v(E-E_\nu)}
\delta(N-N_\nu)~\delta(E-E_\nu),
\end{array}
\label{rho1}
\end{equation}
where the sum runs over all possible configurations $\nu$, each of them specified by the total energy $E_\nu$ and number of particles $N_\nu$. The two exponential functions inserted in the last equality, although irrelevant here, will help us to prevent possible nonphysical divergences. 

Using an integral representation for the delta functions, Eq.~(\ref{rho1}) becomes
\begin{equation}
\rho_{\mathrm{mb}}(E,N)=
\frac{1}{(2\pi i)^2} 
\int_{C_\beta} d\beta 
\int_{C_\alpha} d\alpha 
~e^{s(\beta,\alpha)}, 
\label{rho2x}
\end{equation}
where
\begin{equation}
s(\beta,\alpha)= -\beta 
[\omega(\beta,\alpha)-E] +\alpha N .
\label{S}
\end{equation}
The path of integration $C_\beta$ is a straight line that runs from $v-i\infty$ to $v+i\infty$, while $C_\alpha$ runs from $u-i\infty$ to $u+i\infty$. We have also defined the new variables
\begin{equation}
\beta=v+ia  
\quad \mathrm{and}\quad 
\alpha=u+ib,
\end{equation}
and the functions
\begin{equation}
\omega(\beta,\alpha) = 
-\frac{1}{\beta} \ln Z(\beta,\alpha)
\quad\mathrm{and}\quad
Z(\beta,\alpha)= \sum_\nu 
e^{-\beta E_\nu - \alpha N_\nu}.
\label{GCdef1}
\end{equation}
In Eqs.~(\ref{rho1}) and (\ref{rho2x}), $u$ and $v$ are arbitrary real numbers. They can be conveniently chosen by demanding the convergence of the sum in the last expression. In the present approximation, for each configuration $\nu$, the term $e^{-\beta E_\nu - \alpha N_\nu}$ may be written in terms of the single-particle energy levels $\epsilon_i$ of the mean-field potential, and their respective numbers of occupation. By counting each possible configuration explicitly, we find
\begin{equation}
Z(\beta,\alpha) =
\prod_{i=0}^{\infty}
(1- e^{-(\alpha+\beta \epsilon_i)})^{-1}
\label{partition}
\end{equation}
and, therefore,
\begin{equation}
\begin{array}{lll}
\omega(\beta,\alpha) &=& \displaystyle
\frac{1}{\beta} \sum_{i=0}^{\infty} 
\ln(1- e^{-(\alpha+\beta \epsilon_i)})=
\frac{1}{\beta} \int_{\epsilon_0}^{\infty} 
\rho_{\mathrm{sp}}(\epsilon)
\ln(1- e^{-(\alpha+\beta \epsilon)})~d\epsilon, 
\end{array}
\label{omega}
\end{equation}
where the single-particle density of states $\rho_{\mathrm{sp}}(\epsilon)$ is the following sum of delta functions,
\begin{equation}
\rho_{\mathrm{sp}}(\epsilon)=
\sum_{i=0}^{\infty}
\delta(\epsilon-\epsilon_i).
\end{equation}
Equations~(\ref{partition}) and~(\ref{omega}) were obtained by assuming that $|e^{-(\alpha+\beta\epsilon_i)}|<1$. Considering, without lost of generality, that $\epsilon_0=0$, one concludes that this inequality is satisfied by imposing: $u>0$ and $v>-u/\epsilon_i$. If one also considers the case where $\epsilon_i\rightarrow\infty$ (for some $i$), one gets that both $u$ and $v$ should be positive numbers. 

At last, we show derivatives of Eq.~{(\ref{omega})} that will be useful later, 
\begin{equation}
\mathcal{N}(\beta,\alpha) \equiv
\frac{\partial[\beta\omega]}{\partial\alpha} =
\int_{\epsilon_0}^{\infty}
\frac{\rho_{\mathrm{sp}}(\epsilon)}
{e^{\alpha+\beta \epsilon}- 1}~d\epsilon
\label{n}
\end{equation}
and
\begin{equation}
\mathcal{E}(\beta,\alpha)\equiv
\frac{\partial[\beta\omega]}{\partial\beta} =
\int_{\epsilon_0}^{\infty}
\frac{\epsilon~\rho_{\mathrm{sp}}(\epsilon)}
{e^{\alpha+\beta \epsilon}-1}~d\epsilon.
\label{e}
\end{equation}

A comparison of Eqs.~(\ref{S}), (\ref{omega}), (\ref{n}) and (\ref{e}) with thermodynamical functions suggests that we can identify $\omega$, $\cal N$, $\cal E$ and $s$ as, respectively, the grand-canonical potential, mean number of particles, mean energy, and entropy. It can be done provided that $\beta$ and $\alpha$ be identified as $1/(k_BT)$ and $-\mu/(k_BT)$, respectively, where $k_B$ is the Boltzmann constant, $T$ the temperature, and $\mu$ the chemical potential. We point out that such identification seems to be in conflict with the fact that, in the present formalism, $\beta$ and $\alpha$ are not constant; they are integration variables of Eq.~(\ref{rho2x}). However, as we will see in the following, in the thermodynamical limit, the region of the $(\beta, \alpha)$-complex space relevant to evaluate the line integral~(\ref{rho2x}) is the vicinity of the saddle point $(\beta_0,\alpha_0)$ of the integrand. Thus, to claim the aforementioned correspondence, functions should be evaluated at $(\beta_0,\alpha_0)=(1/[k_BT],-\mu/[k_BT])$.

Formally, the integral representation of $\rho_{\rm mb}$ is given by Eqs.~(\ref{rho2x}) and~(\ref{omega}), which establish the connection between $\rho_{\rm mb}$ and $\rho_{\rm sp}$. Solving integral~(\ref{rho2x}) will be the goal of the next sections, where new approximating assumptions will be used. 

\section{Thermodynamical limit and the saddle point method}
\label{spm}

We now return to integral~(\ref{rho2x}), which will be performed in the thermodynamical limit using the saddle point (or steepest descent) method~\cite{bleistein}. Formally, such a limit is achieved by supposing that $|s(\beta,\alpha)/\lambda|\sim 1$, where $\lambda$ is a large positive real number, and the method is asymptotically accurate as $\lambda\to\infty$. Physically, the limit arises when $E$ and $N$ are large quantities. In practice, under this condition the complex argument $s$ produces very rapid oscillations in the integrand of Eq.~(\ref{rho2x}) along any generic path in the $(\beta,\alpha)$-complex space, so that evaluating the integral with these curves becomes an impossible task. Interestingly, however, integrating along the steepest descent paths emerging from a saddle point is possible, as stated by the method. Following its prescription, if one can exclude contributions from other potential critical points, as extremals of the integration curve or non-analytical points, the relevant contribution to the path integral~(\ref{rho2x}) comes exclusively from the vicinity of the saddle point $(\beta_0,\alpha_0)$ defined by
\begin{equation}
\mathcal{E}(\beta_0,\alpha_0)=E
\quad\mathrm{and}\quad
\mathcal{N}(\beta_0,\alpha_0)=N.
\label{spcs}
\end{equation}
Then, expanding Eq.~(\ref{rho2x}) up to second order around the saddle point produces
\begin{equation}
\rho^{\so}_{\mathrm{mb}}(E,N) =
\frac{e^{{s}(\beta_0,\alpha_0)}}{(2\pi i)^2} 
\int_{C'} d\beta~ d\alpha 
~e^{\frac{1}{2} \delta^2 {s}}, 
\label{rhoexp}
\end{equation}
where 
\begin{equation}
\delta^2 {s}  =
\left(\begin{array}{cc}
\delta\alpha &\delta\beta
\end{array}\right)
\mathrm{D}(\beta_0,\alpha_0)
\left(\begin{array}{c}
\delta\alpha \\ \delta\beta
\end{array}\right),
\label{delta2S}
\end{equation}
with $\delta\alpha=(\alpha-\alpha_0)$, $\delta\beta=(\beta-\beta_0)$, and 
\begin{equation}
\mathrm{D}(\beta_0,\alpha_0) =
\left.\left(\begin{array}{cc}
\frac{\partial^2 s}{\partial\alpha^2 }
&\frac{\partial^2  s}{\partial\alpha \partial\beta}\\
\frac{\partial^2  s}{\partial\beta \partial\alpha}
&\frac{\partial^2  s}{\partial\beta^2}
\end{array}\right)\right|_{(\beta_0,\alpha_0)}.
\end{equation}
The new path of integration $C'$ arises from the deformation of the original one, procedure that should be done in order to include the saddle point $(\beta_0,\alpha_0)$ in the contour of integration. Besides, in the vicinity of this point, $C'$ must coincide with the steepest descent path passing through $(\beta_0,\alpha_0)$. A rigorous application of the method demands a justification of the deformation, which may be done by using the Cauchy's Integral Theorem. However, given the difficulties to accomplish this task in a general scenario, as usual, we just assume that this step can be performed. 

By evaluating the gaussian integral, Eq.~(\ref{rhoexp}) becomes
\begin{equation}
\rho^{\so}_{\mathrm{mb}}(E,N) =
\frac{1}{2\pi} 
\frac{ e^{s(\beta_0,\alpha_0)}}
{\sqrt{|\det \mathrm{D}(\beta_0,\alpha_0)|}} ,
\label{rhof}
\end{equation}
which establishes how the density of states depends on the two parameters, $\beta_0$ and $\alpha_0$, whose relation with $E$ and $N$ is given by Eq.~(\ref{spcs}). We remind that the subject presented in the present and previous sections can be found in the literature~\cite{mbdsB1,mbdsB2}, as well as its analogue for fermions~\cite{mbdsF1,mbdsF2}.

\section{Uniform formula}
\label{un}

An important requirement needed to the above application of the steepest descent method is the assumption that the considered saddle point be isolated from other ones and singularities as well. Indeed, it is assumed that there is just one saddle point in the present work so that we do not have to deal with the potential problem of their coalescence. On the other hand, when we write the integrand of Eq.~(\ref{rho2x}) as 
\begin{equation}
e^{s(\beta,\alpha)}
= \frac{e^{\beta E + \alpha N}}
{\prod_{i=0}^{\infty}
\left(1-e^{-(\alpha+\beta\epsilon_i)}\right) },
\end{equation}
we realize that it diverges whenever $\alpha+\beta\epsilon_i = \pm 2n\pi i$, for $n=0,1,2,\ldots$. In particular, since that $\epsilon_0=0$, when $\alpha_0\rightarrow0$, independently of the value of $\beta_0$, Eq.~(\ref{rhof}) fails due to the influence of the non-analyticity at $\alpha=0$.  In the present section we perform a uniform approximation on $\rho_{\rm mb}$, seeking to avoid {\em just this source} of problem. In this sense, we will assume that the saddle point $(\tilde\beta_0, \tilde\alpha_0)$ of the new approach is such that the function $[1-e^{-(\tilde\alpha_0+\tilde\beta_0\epsilon_i)}]$ does not vanish for every excited state~$\epsilon_i$. This is the simplest improvement we can do. Summarizing, as the saddle point $\alpha_0$ may lie near the amplitude critical point $\alpha=0$, the second order approximation is said to be non-uniform with respect to $\alpha_0$, giving rise to inaccurate results when $\alpha_0\to 0$. To avoid it, a uniform approximation~\cite{bleistein} can be useful, which will be developed in the following. It is important to mention that Holthaus and Kalinowski~\cite{holthausANN99} considered similar techniques to evaluate other statistical functions for many-boson systems, also taking into account the effects of this singularity.

Since the approximation~(\ref{rhof}) breaks down when the ground state starts to be sufficiently occupied [see Eqs~(\ref{n}) and~(\ref{e})], we will give it a special treatment, which consists in writing the integrand of Eq.~(\ref{rho2x}) as 
\begin{equation}
e^{s(\beta,\alpha)} =
\frac{e^{s_e(\beta,\alpha)}}{1-e^{-\alpha}} ,
\end{equation}
where
\begin{equation}
s_e(\beta,\alpha)=-\beta [\omega_e(\beta,\alpha)-E] +\alpha N 
\label{Se}
\end{equation}
and $\omega_e(\beta,\alpha)$ is given by Eq.~(\ref{omega}), replacing $\rho_{\mathrm{sp}}(\epsilon)$ by
\begin{equation}
\rho_{\mathrm{sp}}^{(e)}(\epsilon) \equiv \sum_{i=1}^{\infty}
\delta(\epsilon-\epsilon_i) .
\label{exc}
\end{equation}

In terms of $s_e(\beta,\alpha)$, Eq.~(\ref{rho2x}) becomes
\begin{equation}
\begin{array}{lll}
\rho_{\mathrm{mb}}(E,N) &=& \displaystyle
\frac{1}{(2\pi i)^2} 
\int_{C_\beta} 
\int_{C_\alpha}d\beta~d\alpha
~\frac{e^{s_e(\beta,\alpha)}}
{1-e^{-\alpha}} =
\frac{1}{2\pi i} 
\int_{C_\alpha}d\alpha
~\frac{I_\beta(\alpha)}
{1-e^{-\alpha}} ,
\end{array}
\label{mainI}
\end{equation}
where we define
\begin{equation}
I_\beta(\alpha) =\frac{1}{2\pi i} 
\int_{C_\beta} 
d\beta~e^{s_e(\beta,\alpha)}.
\label{Ibeta}
\end{equation}

\subsection{Integration on $\beta$}

Integral $I_\beta$ can be easily performed by means of the second order steepest descent method, provided that we take the thermodynamical limit in the same way as discussed in the beginning of Sect.~\ref{spm}. {\it A priori,} the saddle point $\tilde\beta_\alpha$ of integral~(\ref{Ibeta}) needs to be sufficiently far from poles of the integrand, namely, $|1-e^{-(\alpha+\tilde\beta_\alpha\epsilon_i)}|$ cannot vanish for any value of $\alpha$ belonging to $C_\alpha$. The saddle point $\tilde\beta_\alpha=\tilde\beta_\alpha(\alpha)$ is given by
\begin{equation}
\left.\left(
\frac{\partial s_e}{\partial \beta}
\right)\right|_{\beta=\tilde\beta_\alpha} = 0
\quad\Longrightarrow\quad
E=\mathcal{E}_e(\tilde\beta_\alpha,\alpha),
\end{equation}
where
\begin{equation}
\mathcal{E}_e(\beta,\alpha) \equiv
\int_{\epsilon_1}^{\infty}
\frac{\epsilon~\rho_{\mathrm{sp}}^{(e)}(\epsilon)}
{e^{\alpha+\beta \epsilon}-1}
~d\epsilon.
\end{equation}
Considering in addition that the function $ \left.\left( \partial^2s_e/\partial\beta^2 \right)\right|_{\beta=\tilde\beta_\alpha}$ does not vanish, and assuming that the original path of integration can be deformed onto steepest descent paths of $\tilde\beta_\alpha$, we easily find
\begin{eqnarray}
I_\beta(\alpha) \approx
I_\beta^{(2)}(\alpha) \equiv 
\left[2\pi\left.\left(
\partial^2s_e/\partial\beta^2
\right)\right|_{\beta=\tilde\beta_\alpha}\right]^{-1/2}
\exp\left\{s_e(\tilde\beta_\alpha,\alpha)\right\}.
\label{Ibetasolved}
\end{eqnarray}

For the next calculations, it will be useful to define
\begin{equation}
\sigma(\alpha)\equiv s_e(\tilde\beta_\alpha[\alpha],\alpha)-
\frac12 \ln \left[ 
\left.\left(
\frac{\partial^2s_e}{\partial\beta^2}
\right)\right|_{(\tilde\beta_\alpha[\alpha],\alpha)}\right],
\label{def1}
\end{equation}
so that Eq.~(\ref{mainI}) becomes
\begin{eqnarray}
\rho_{\mathrm{mb}}(E,N) \approx \frac{1}{(2\pi )^{3/2}i} 
\int_{C_\alpha}d\alpha
~\frac{e^{\sigma(\alpha)}}
{1-e^{-\alpha}}  .
\label{mainI3}
\end{eqnarray}

\subsection{Integration on $\alpha$ -- implicit change of variable}

To solve integral~(\ref{mainI3}), we first look for the saddle point $\tilde\alpha_0$ of the exponent $\sigma(\alpha)$,
\begin{equation}
\left.\left(
\frac{\partial\sigma}{\partial\alpha}
\right)\right|_{\alpha=\tilde\alpha_0}
= 0.
\end{equation}
It is easy to show that the pair $(\tilde\beta_0\equiv\tilde\beta_\alpha[\tilde\alpha_0],\tilde\alpha_0)$, up to the leading order in $\lambda$, is given by 
\begin{equation}
E = \mathcal{E}_e(\tilde\beta_0,\tilde\alpha_0)
\quad\mathrm{and}\quad
N = \mathcal{N}_e(\tilde\beta_0,\tilde\alpha_0),
\label{cp1}
\end{equation}
with
\begin{equation}
\mathcal{N}_e(\beta,\alpha) \equiv
\int_{\epsilon_1}^{\infty}
\frac{\rho_{\mathrm{sp}}^{(e)}(\epsilon)}
{e^{\alpha+\beta \epsilon}-1}
~d\epsilon .
\end{equation}
Within this new approach, we point out that $\tilde\beta_0$ and $\tilde\alpha_0$ can no longer be directly related to the inverse of temperature and chemical potential, respectively, except when the population of the ground state can be disregarded in comparison with~$N$, namely, when $e^{-\tilde\alpha_0}$ is sufficiently small.

Equation~(\ref{mainI3}) clearly shows that the application of the saddle point method fails when $\tilde\alpha_0\to 0$. In order to overcome this problem by means of a uniform formula, we {\em implicitly} define a change of variable $\alpha=\alpha(t)$, with inverse $t=t(\alpha)$, through the map
\begin{equation}
\sigma(\alpha) = \phi (t) ,
\quad\mathrm{with}\quad
\phi (t) \equiv \lambda \left(
-\frac{t^2}{2}-\gamma t + \chi \right).
\label{map}
\end{equation}
The saddle point of $\phi (t)$ is $t_0 =-\gamma$. The first condition for a proper mapping is the equivalence of saddle points, namely, $\alpha(t_0)=\tilde\alpha_0$, which produces
\begin{equation}
\sigma(\tilde\alpha_0) = \phi (t_0)
\quad\Longrightarrow \quad
\sigma(\tilde\alpha_0) = \lambda 
\left(\frac{\gamma^2}{2} + \chi \right) . 
\end{equation}
The second convenient condition is mapping the pole $\alpha=0$ onto the origin $t=0$, namely, $t(0)=0$. Thus,
\begin{equation}
\sigma(0) = \phi (0)
\quad\Longrightarrow \quad
\sigma(0) = \lambda \chi .  
\label{cond2}
\end{equation}
Therefore, $\gamma$ and $\chi$, in terms of the original function $\sigma(\alpha)$, are given by
\begin{eqnarray}
\sqrt{\lambda}\gamma = 
\pm \left\{2[\sigma(\tilde\alpha_0) 
- \sigma(0)]\right\}^{1/2}\equiv \pm 
\left[2\Delta\sigma\right]^{1/2}
~\mathrm{and}~~
\lambda \chi  = \sigma(0) . 
\label{Dsigma}
\end{eqnarray}

There is still one point to solve: which branch on the definition of $\gamma$ should be chosen? To answer this question, we should solve Eq.~(\ref{map}) finding an expression for $t(\alpha)$. Applying this expression to the point $\alpha=0$, we conclude that $\gamma=+\sqrt{\gamma^2}$, otherwise we will not find $t(0)=0$.

Then, by changing variables, integral~(\ref{mainI3}) becomes
\begin{eqnarray}
\rho_{\mathrm{mb}}(E,N) \approx \frac{1}{(2\pi )^{3/2}i} 
\int_{C_t}
\left(\frac{e^{\phi(t)}}{t}\right) ~
G(t)~dt ,
\label{rhoun}
\end{eqnarray}
where $C_t$ is the image of $C_\alpha$ under the implicit transformation~(\ref{map}), and
\begin{equation}
G(t) = 
\frac{t}{1-e^{-\alpha(t)}} 
\frac{d\alpha}{dt}.
\label{jacobG}
\end{equation}
To give continuity and evaluate Eq.~(\ref{rhoun}), we need an explicit formula for $G(t)$. At this point, we conveniently write
\begin{equation}
G(t) = 
G(0) + A t  + t(t+\gamma)H_0(t),
\label{Gdef}
\end{equation}
where
\begin{equation}
A=\sqrt\lambda\left(
\frac{G(0)  - G(-\gamma) }{\sqrt\lambda\gamma}
\right)\equiv \sqrt\lambda~\bar A.
\label{Adef}
\end{equation}
Functions $G(0)$ and $G(t_0)=G(-\gamma)$ are deduced in the Appendix~\ref{app1}:
\begin{equation}
G(0)=1
\quad\mathrm{and}\quad
G(t_0)= -i
\frac{[2\Delta\sigma]^{1/2}
[\sigma^{(2)}(\tilde\alpha_0)]^{-1/2}}{1-e^{-\tilde\alpha_0}},
\end{equation}
where we have defined $\sigma^{(2)}(\alpha)\equiv d^2\sigma/d\alpha^2 $. The third term of Eq.~(\ref{Gdef}), which contains $H_0(t)$, will be disregarded~\cite{bleistein} because, by replacing it on Eq.~(\ref{rhoun}), it gives origin to terms of order $\lambda^{-1}$ in comparison with the result that we will derive here [Eq.~(\ref{ov}) below]. Thus, defining
\begin{equation}
W_{r}^{(C_T)} (z)\equiv \int_{C_T} T^r 
\exp\left\{ -\frac{T^2}{2} - z T\right\}dT ,
\end{equation}
we have
\begin{equation}
\begin{array}{lll}
\rho_{\mathrm{mb}}^{\mathrm{un}}(E,N) &=& 
\displaystyle \frac{e^{\sigma(0)}}{(2\pi )^{3/2}i}
~W_{-1}^{(C_T)}\left([2\Delta\sigma]^{1/2}\right) +
\frac{\bar A~e^{\sigma(0)}}{(2\pi )^{3/2}i}~
W_{0}^{(C_T)}\left([2\Delta\sigma]^{1/2}\right),
\label{rhoMBun1}
\end{array}
\end{equation}
where $C_T$ is the image of $C_t$ under the transformation $t=T/\sqrt{\lambda}$, meaning that $C_T$ is the same as $C_t$ except for a scaling factor. By its turn,  $C_t$, as already said, is the image of $C_\alpha$ under transformation~(\ref{map}). According to Bleistein and Handelsman~\cite{bleistein}, there are only two independent choices for $C_T$. One path is a loop around the origin, while the other does not encircle it. As our original contour of integration is a straight line, namely, it does not encircle the origin, we assume that $C_T$ is also so. In this case, we have
\begin{equation}
\begin{array}{lll}
\displaystyle W_{ 0}^{(C_T)} (z) =  
\sqrt{2\pi}~ e^{z^2/2} &
\mathrm{and}&
\displaystyle 
W_{-1}^{(C_T)} (z) = 
-i\pi ~\mathrm{erfc}\left[ \frac{-iz}{\sqrt2}\right],
\end{array}
\end{equation}
where $\mathrm{erfc}(z)$ is the complementary error function
\begin{equation}
\mathrm{erfc} (z)= \frac{2}{\sqrt\pi}\int_{z}^{\infty} 
e^{-\zeta^2} d\zeta =
1 - \frac{2}{\sqrt\pi}\sum_{k=0}^{k=\infty}
\frac{(-1)^{k}z^{2k+1}}{k!(2k+1)} .
\label{err}
\end{equation}
Using these last results on Eq.~(\ref{rhoMBun1}), we finally have
\begin{equation}
\begin{array}{lll}
\rho_{\mathrm{mb}}^{\mathrm{un}}(E,N) &=& 
\displaystyle 
-\frac{e^{\sigma(0)}}{2\sqrt{2\pi}}~
\mathrm{erfc}\left(-i\sqrt{\Delta\sigma}\right)
-\frac{\bar A~e^{\sigma(\tilde\alpha_0)}}{2\pi i}.
\end{array}
\label{ov}
\end{equation}
This expression, in principle, should be uniformly good for all values of $\tilde\alpha_0$, and its relative error is of order $\lambda^{-1}$. It is worth to remind that $\bar A$ is of order $\lambda^{-1/2}$ while the pre-exponential factor present in the first term is independent of $\lambda$. Equation~(\ref{ov}) and its consequent expressions constitute the first contribution of the present work. In the following, we will discuss two limits: $\tilde\alpha_0 \to0$, the regime which our approach intends to improve; and $\tilde\alpha_0 \gg 1$, where Eq.~(\ref{ov}) is expected to approach Eq.~(\ref{rhof}).

\subsection{Limit $\tilde\alpha_0 \to0$ }

A simpler expression for the uniform formula~(\ref{ov}) can be achieved by considering the formal limit $\tilde\alpha_0 \to 0$. In order to do that, previous results should be derived. First, according to Eq.~(\ref{exp1}), we conclude that $\Delta \sigma$ is of order $\tilde\alpha_0^2$:
\begin{equation}
\Delta \sigma = -\frac12 \sigma^{(2)}_0 \tilde \alpha_0^2
\left[
1+\frac{2}{3} \frac{\sigma^{(3)}_0}{\sigma^{(2)}_0}\tilde\alpha_0
+\frac{1}{4} \frac{\sigma^{(4)}_0}{\sigma^{(2)}_0}\tilde\alpha_0^2
+\mathcal{O}(\tilde\alpha_0^3)
\right],
\label{Dsigma0}
\end{equation}
where $\sigma^{(k)}_0\equiv (d^k\sigma/d\alpha^k)|_{\alpha=0}$. In this section we will face some functions like this, which may involve powers of the large number $\lambda$ times powers of $\tilde\alpha_0$. Then, we precisely define the limit $\tilde\alpha_0 \to 0$ by additionally assuming that $\lambda\,\tilde\alpha_0 = \mathcal{O}(\tilde\alpha_0)$.

By combining Eq.~(\ref{Dsigma0}) with the expansion
\begin{equation}
\begin{array}{rll}
\displaystyle
\left.\left(\frac{d^2\sigma}{d\alpha^2}
\right)\right|_{\alpha=\tilde\alpha_0}&=&
\displaystyle
\sigma^{(2)}_0
\left[
1+\frac{\sigma^{(3)}_0}{\sigma^{(2)}_0}\tilde\alpha_0+
\frac12 \frac{\sigma^{(4)}_0}{\sigma^{(2)}_0}\tilde\alpha_0^2
+ \mathcal{O}(\tilde\alpha_0^3)
\right],
\end{array}
\end{equation}
we find
\begin{equation}
\begin{array}{rll}
\displaystyle
\bar A &=&
\displaystyle
\frac{1-G(t_0)}{\sqrt{2\Delta\sigma}} = 
\frac{\mathcal A + \mathcal B \tilde\alpha_0 + 
\mathcal{O}(\tilde\alpha_0^2)}
{[\sigma^{(2)}_0]^{1/2}},
\end{array}
\end{equation}
where $\mathcal A$ and $\mathcal B$ are explicitly given by
\begin{equation}
\begin{array}{lll}
\displaystyle
\mathcal A =
\frac{i}{2} \left(
1 - \frac13\frac{\sigma^{(3)}_0}{\sigma^{(2)}_0} 
\right)
&\mathrm{and}&
 \displaystyle\mathcal B = 
\frac{i}{12}
\left[
1-3\left(
\frac12\frac{\sigma^{(4)}_0}{\sigma^{(2)}_0} +
\frac{\sigma^{(3)}_0}{\sigma^{(2)}_0} 
\right)+\frac52
\left(\frac{\sigma^{(3)}_0}{\sigma^{(2)}_0}
\right)^2\right].
\end{array}
\label{ABl}
\end{equation}
Thus, according to expansion~(\ref{err}) we write
\begin{equation}
\begin{array}{rll}
\displaystyle\mathrm{erf}\left(-i\sqrt{\Delta \sigma}\right)&=&
\displaystyle
1- \sqrt{\frac{2}{\pi}} \,[\sigma^{(2)}_0]^{1/2}\tilde\alpha_0
\left[1+ \mathcal{O}(\tilde\alpha_0)\right]
+\mathcal{O}\left(\tilde\alpha_0^3\right),
\end{array}
\end{equation}
and finally
\begin{equation}
\begin{array}{lll}
\rho_{\mathrm{mb}}^{\mathrm{un}}(E,N) &=&
\displaystyle 
-\frac{e^{\sigma(0)}}{2\sqrt{2\pi}}
\left\{
\left[1-i\sqrt{\frac{2}{\pi}}\frac{\mathcal A}{[\sigma^{(2)}_0]^{1/2}} \right]
-\sqrt{\frac{2}{\pi}}
\left[
\frac{\sigma^{(2)}_0 + i\mathcal B}{[\sigma^{(2)}_0]^{1/2}}\right]\tilde\alpha_0 
+\mathcal{O}\left(\tilde\alpha_0^2\right)\right\}.
\end{array}
\label{rhoqdlp}
\end{equation}
If one just keeps the correction terms on $\tilde\alpha_0$ up to leading order, we find
\begin{equation}
\rho_{\mathrm{mb}}^{\mathrm{un}}(E,N;\tilde\alpha_0\to 0) =\displaystyle 
-\frac{\mathcal C e^{\sigma(0)}}{2\sqrt{2\pi}},
\label{rhoqdl}
\end{equation}
whose relative error is of order $ \tilde\alpha_0^2$. Here, we have also defined $\mathcal C = c_0\, e^{-(c_1/c_0)\tilde\alpha_0}$, which exclusively contains the dependence on $\tilde\alpha_0$, and
\begin{equation}
c_0=1-i\sqrt{\frac{2}{\pi}}\frac{\mathcal A}{[\sigma^{(2)}_0]^{1/2}}
\quad\mathrm{and}\quad
c_1= \sqrt{\frac{2}{\pi}}\frac{\sigma^{(2)}_0 + i\mathcal B}{[\sigma^{(2)}_0]^{1/2}}.
\end{equation}

\subsection{Limit $\tilde\alpha_0 \gg1$}

In contrast to the last section, here we will consider the case where $\tilde\alpha_0$ is large. In this limit, to find an expression for Eq.~(\ref{ov}), it is equivalent and easier to formally consider the limit $|\Delta \sigma| \to\infty$. As well as in the previous section, it is important to clarify the role of $\lambda$ in our manipulations; we will assume that $\lambda/\Delta\sigma=\mathcal{O}(\Delta\sigma^{-1})$. A further assumption adopted here, which will be proved true in the next section, is considering $\Delta \sigma=|\Delta \sigma|$. 

We start by presenting the following expansion, valid when $|z|\rightarrow\infty$,
\begin{equation}
\mathrm{erfc}(z)= 
\left[ 1 - \frac{\sqrt{z^2}}{z} \right] + 
\frac{e^{-z^2}}{\sqrt\pi z}
\left[1 + \mathcal{O}\left(\frac{1}{z^2}\right) 
\right] .
\end{equation}
Using it, we find
\begin{equation}
\begin{array}{lll}
\displaystyle
\mathrm{erfc}\left(-i\sqrt{\Delta \sigma}\right)
&=& 
\displaystyle
\frac{i\,e^{-\Delta \sigma}}{\sqrt{\pi \Delta \sigma}}
\left[1 + \mathcal{O}\left(\frac{1}{\Delta \sigma}
\right) 
\right],
\end{array}
\end{equation}
which, together with
\begin{equation}
\bar A =
\frac{1}{[2\Delta \sigma]^{1/2}}
+i  \frac{[\sigma^{(2)}(\tilde\alpha_0)]^{-1/2}}{1-e^{-\tilde\alpha_0}}
\end{equation}
produces
\begin{equation}
\begin{array}{l}
\rho_{\mathrm{mb}}^{\mathrm{un}}(E,N) =
\displaystyle -
\frac{e^{\sigma(\tilde\alpha_0)}}{2\pi}
\left\{ 
\frac{[\sigma^{(2)}(\tilde\alpha_0)]^{-1/2}}{1-e^{-\tilde\alpha_0}}
+\mathcal{O}\left(\frac{1}{\sqrt{\Delta\sigma}}\right)
\right\}.
\end{array}
\label{newtooldmb}
\end{equation}
Finally, by keeping only the leading order we find
\begin{equation}
\rho_{\mathrm{mb}}^{\mathrm{un}}(E,N;\tilde\alpha_0\gg1) = -
\displaystyle 
\frac{e^{\sigma(\tilde\alpha_0)}}{2\pi[\sigma^{(2)}
(\tilde\alpha_0)]^{1/2}},
\label{newtooldmb2}
\end{equation}
whose relative error is of order $\Delta\sigma^{-1}$.

\subsection{Back to original variables}
\label{bov}

In order to discuss the last results [Eqs.~(\ref{ov}), (\ref{rhoqdl}), and~(\ref{newtooldmb2})], it is better to write them using the original variables. We then define $\tilde\beta_0^*$ according to $\mathcal E_e (\tilde\beta_0^*,0)=E$, so that 
\begin{equation}
\begin{array}{lll}
\displaystyle\sigma(0) =
 s_e(\tilde\beta_0^*,0)-\frac12 \ln\left[
\left. \frac{\partial^2 s_e}
{\partial\beta^2}\right|_{(\tilde\beta_0^*,0)}
\right]
&\mathrm{and}&
\displaystyle
\bar A =
\frac{1}{[2\Delta s_e]^{1/2}}
+i \frac{ 
[S_e^{(2)}]^{-1/2}}{1-e^{-\tilde\alpha_0}},
\end{array}
\end{equation}
where we defined
\begin{equation}
\begin{array}{l}
\displaystyle \Delta s_e\equiv\displaystyle 
s_e(\tilde\beta_0,\tilde\alpha_0) - s_e(\tilde\beta_0^*,0)
-\frac12 \ln\left[\left(\left. \frac{\partial^2 s_e}
{\partial\beta^2}\right|_{(\tilde\beta_0,\tilde\alpha_0)}\right)
\left(\left. \frac{\partial^2 s_e}
{\partial\beta^2}\right|_{(\tilde\beta_0^*,0)}\right)^{-1}
\right], \\\\\displaystyle
\sigma^{(2)}(\tilde\alpha_0)=S_e^{(2)} \equiv\displaystyle 
\left.\left\{
\left(\frac{\partial^2s_e}{\partial \beta^2}\right)^{-1}
\left[
\frac{\partial^2s_e}{\partial\alpha\partial\beta}\frac{\partial^2s_e}{\partial\alpha\partial\beta}-
\frac{\partial^2s_e}{\partial\alpha^2}\frac{\partial^2s_e}{\partial \beta^2}\right] 
\right\}\right|_{(\tilde\beta_0,\tilde\alpha_0)}.
\end{array}
\label{newtoold}
\end{equation}
Notice that the connection between $\sigma^{(2)} (\tilde\alpha_0)$ and $S_e^{(2)} $ has been established in Appendix~\ref{app3}, Eq.~(\ref{C6}). We finally write Eqs.~(\ref{ov}), (\ref{rhoqdl}), and~(\ref{newtooldmb2}) as
\begin{equation}
\begin{array}{lll}
\rho_{\mathrm{mb}}^{\mathrm{un}}(E,N) &=&
\displaystyle 
\left|\left[ 
\left.\left( \frac{\partial^2 s_e}
{\partial\beta^2}\right)\right|_{(\tilde\beta_0^*,0)}
\right]^{-\frac12}
\frac{e^{s_e(\tilde\beta_0^*,0)}}{2\sqrt{2\pi}}
\left\{
\mathrm{erfc}\left(-i\sqrt{\Delta s_e}\right)
+i\bar{A}\sqrt{\frac{2}{\pi}}~e^{\Delta s_e}
\right\}\right|,\\\\
\rho_{\mathrm{mb}}^{\mathrm{un}}(E,N;\tilde\alpha_0\to0) &=&
\displaystyle 
\frac{\mathcal C\, e^{ s_e(\tilde\beta_0^*,0)}}{2\sqrt{2\pi}}
\left|\left. \frac{\partial^2 s_e}
{\partial\beta^2}\right|_{(\tilde\beta_0^*,0)}
\right|^{-1/2},\\\\
\rho_{\mathrm{mb}}^{\mathrm{un}}(E,N;\tilde\alpha_0\gg1) &=&
\displaystyle 
\frac{e^{s_e(\tilde\beta_0,\tilde\alpha_0)}}{2\pi}
\left|\left.
\frac{\partial^2s_e}{\partial\alpha\partial\beta}\frac{\partial^2s_e}{\partial\alpha\partial\beta}-
\frac{\partial^2s_e}{\partial\alpha^2}\frac{\partial^2s_e}{\partial \beta^2} 
\right|_{(\tilde\beta_0,\tilde\alpha_0)}\right|^{-1/2},
\end{array}
\label{unset}
\end{equation}
where the modulus operation was arbitrarily used to assure the physical meaning of the expressions; notice that a minus sign could simply arise by performing the integration with an inverse path, situation that is not controlled by the present general application of the method. The third equation is the same as Eq.~(\ref{rhof}), except for the definition of the saddle points and the fact that $s$ was replaced by $s_e$. However, the differences vanish in the adopted limit, because $\tilde\alpha_0 \gg 1$ implies $s_e(\beta,\alpha) \to s(\beta,\alpha)$. This conclusion is expected and validates the assumptions that we used in the last section, since that, in the considered limit, the saddle point is far from the singularity, condition required to derive Eq.(\ref{rhof}). Concerning the term $\mathcal C$ in the expression of $\rho_{\mathrm{mb}}^{\mathrm{un}}(E,N;\tilde\alpha_0\to0)$, we did not rewrite it in terms of the original variables due to the difficulties to deal with it. We will return to this point opportunely.

In the next section, we will perform a first needed step to submit $\rho_{\mathrm{mb}}^{\mathrm{un}}(E,N;\tilde\alpha_0\to0)$ to tests, which consists in using semiclassical methods to introduce in the theory information about the single-particle spectrum, and, consequently, have an explicit formula for $s$ and $s_e$.

\section{Smooth Approximation}
\label{smo}

The quantities $s$ and $s_e$ are the main functions for the calculation of the expressions $\rho_{\mathrm{mb}}^{\so}$ and $\rho_{\mathrm{mb}}^{\rm un}$, respectively. However, to evaluate them, we still need to know the single-particle spectrum.  To provide such information, an option is to introduce another approximation in the theory, which is working with the semiclassical single-particle level density $\bar\rho_{\mathrm{sp}}$, instead of the exact one. It can be achieved by means of the Thomas-Fermi method~\cite{brack}, which basically consists in computing the volume of the classically allowed phase space $(\mathbf{q},\mathbf{p})$, measured in unities of the `quantum cell' $(2\pi\hbar)^{f_d}$. We have
\begin{equation}
\bar\rho_{\mathrm{sp}}(\epsilon) = \frac{1}{(2\pi\hbar)^{f_d}} 
\int d^{f_d} q \; d^{f_d} p\; 
\delta[H(\mathbf{q},\mathbf{p})-\epsilon],
\label{smooth}
\end{equation}
where the integral should span the whole phase space. $H(\mathbf{q},\mathbf{p})$ is the Hamiltonian of the equivalent classical problem and $f_d$ is the spatial dimension. The function $\bar\rho_{\mathrm{sp}}(\epsilon)$ is known as the smooth part (or continuous approximation) of $\rho_{\mathrm{sp}}(\epsilon)$, and its inverse can be identified as the mean spacing between two neighboring levels, around $\epsilon$. In general, the last integral leads to power series of $\epsilon$.

\subsection{Second order approximation $\bar{\rho}_{\mathrm{mb}}^{\so}$}

We will assume the simplest form of the semiclassical single-particle spectrum: $ \bar\rho_{\mathrm{sp}}(\epsilon)=c\epsilon^\gamma$, where $c$ is a positive real constant. However, to represent a physical system we must have $\gamma>-1$, because the step function written in the smooth approximation,
\begin{equation}
\bar n_{\mathrm{sp}}(\epsilon) \equiv 
\int^\epsilon \bar \rho_{\mathrm{sp}}(\epsilon')
d\epsilon' = \frac{c}{\gamma+1} \epsilon^{\gamma+1}+1,
\label{smapp}
\end{equation}
has to be monotonically crescent. Notice that a unitary constant is included in the last formula to satisfy $\bar n_{\mathrm{sp}}(0)=1$, which contemplates the ground state $\epsilon_0=0$ in the expression.

Using these ideas, Eq.~(\ref{omega}) becomes
\begin{equation}
\omega(\beta,\alpha) \approx \bar\omega(\beta,\alpha) 
\equiv \frac1\beta
\int_{0}^{\infty} 
\bar \rho_{\mathrm{sp}}(\epsilon) 
\ln(1-e^{-(\alpha+\beta\epsilon)})~d\epsilon .
\label{gcf3}
\end{equation}
As $|e^{-(\alpha+\beta\epsilon)}|<1$, we use geometrical series in order to expand
\begin{equation}
\frac{d}{d\epsilon}\ln(1- e^{-(\alpha+\beta\epsilon)})=
\frac{\beta}{e^{\alpha+\beta\epsilon}-1}=
\beta \sum_{l=1}^{\infty} e^{-l(\alpha+\beta\epsilon)} ,
\label{around0}
\end{equation}
so that integrating Eq.~(\ref{gcf3}) by parts furnishes
\begin{equation}
\bar\omega(\beta,\alpha) = 
-\frac{c\Gamma(\gamma+1) \mathcal{L}_{\gamma+2}(e^{-\alpha})}
{\beta^{\gamma+2}},
\label{betaomegaexp}
\end{equation}
where $\Gamma$ and $\mathcal{L}$ are, respectively, the Gamma and Polylogarithmic functions,
\begin{equation}
\Gamma(\nu+1) = \int_0^\infty x^\nu e^{-x} dx
\quad\mathrm{and}\quad
\mathcal{L}_{\nu}(\chi) = 
\sum_{l=1}^{\infty} \frac{\chi^l}{l^{\nu}}.
\end{equation}
For a fixed $\gamma$, remembering that $|e^{-\alpha}|<1$, we realize that the function $\mathcal{L}_{\gamma+2}(e^{-\alpha})$ has its maximum value at $\alpha\to0$. In such a case, $\mathcal{L}_{\gamma+2}(1)$ becomes the Riemann zeta function $\zeta(\gamma+2)$, which converges for $\gamma>-1$. In addition, the Gamma function $\Gamma(\gamma+1)$ also converges for $\gamma>-1$. 

Function $s(\beta,\alpha)$ [see Eq.~(\ref{S})] can then be approached by
\begin{equation}
\bar s (\beta,\alpha) \equiv  
\frac{c \Gamma(\gamma+1) 
\mathcal{L}_{\gamma+2}(e^{-\alpha})}
{\beta^{\gamma+1}}
+\beta E + \alpha N,
\label{sbar}
\end{equation}
whose derivatives can be directly found,
\begin{equation}
\begin{array}{lll}
\displaystyle{\frac{\partial^2 \bar s}
{\partial \beta^2}} &=& 
\displaystyle{\frac{c (\gamma+2) \Gamma(\gamma+2) 
\mathcal{L}_{\gamma+2}(e^{-\alpha})}{\beta^{\gamma+3}}}, 
\\ \\
\displaystyle{\frac{\partial^2 \bar s}
{\partial \alpha\partial \beta}} &=& 
\displaystyle{ \frac{c \Gamma(\gamma+2) 
\mathcal{L}_{\gamma+1}(e^{-\alpha})}{\beta^{\gamma+2}}
=\frac{\partial^2 \bar s}{\partial \beta\partial \alpha}}, 
\\ \\
\displaystyle{\frac{\partial^2 \bar s}
{\partial \alpha^2}} &=&
\displaystyle{\frac{c \Gamma(\gamma+1) 
\mathcal{L}_{\gamma}(e^{-\alpha})}{\beta^{\gamma+1}}},
\end{array}
\label{Ds}
\end{equation}
where we used the relations
\begin{equation}
\frac{\partial \mathcal{L}_{\nu}(\chi)}{\partial \chi}
= \frac{\mathcal{L}_{\nu-1}(\chi)}{\chi}
\quad\mathrm{and}\quad
\Gamma(\nu+1)=\nu\Gamma(\nu).
\end{equation}
Equations~(\ref{n}) and (\ref{e}) can be analogously evaluated, so that the saddle point $(\beta_0,\alpha_0)$ becomes determined by
\begin{equation}
\begin{array}{lll}
\bar{\mathcal{E}}(\beta_0,\alpha_0) &\equiv& 
\displaystyle \int_{0}^{\infty}
\frac{\epsilon~\bar{\rho}_{\mathrm{sp}}(\epsilon)}
{e^{\alpha_0+\beta_0 \epsilon}-1}~d\epsilon
=\frac{c\Gamma(\gamma+2) \mathcal{L}_{\gamma+2}(e^{-\alpha_0})}
{\beta_0^{\gamma+2}}=E,
\\ \\
\bar{\mathcal{N}}(\beta_0,\alpha_0) &\equiv&\displaystyle
\int_{0}^{\infty}\frac{\bar{\rho}_{\mathrm{sp}}(\epsilon)}
{e^{\alpha_0+\beta_0 \epsilon}- 1}~d\epsilon  =
\frac{c\Gamma(\gamma+1) \mathcal{L}_{\gamma+1}(e^{-\alpha_0})}
{\beta_0^{\gamma+1}}  =N.
\end{array}
\label{gcfTmu}
\end{equation}
By manipulating these equations we can eliminate $\beta_0$ from them so that $\alpha_0$ can be written in terms of the pair $(E,N)$ by means of 
\begin{equation}
\mathcal F_\gamma(\alpha_0) \equiv
\frac{\mathcal{L}_{\gamma+1}(e^{-\alpha_0})}
{[\mathcal{L}_{\gamma+2}(e^{-\alpha_0})]
^{\frac{\gamma+1}{\gamma+2}}} = d,
\quad\mathrm{where}\quad
d\equiv\left( \frac{(\gamma+1)^{\gamma+1}}{c\Gamma(\gamma+1)}
\frac{N^{\gamma+2}}{E^{\gamma+1}}
\right)^{\frac{1}{\gamma+2}}.
\label{alpha0}
\end{equation}
Once Eq.~(\ref{alpha0}) is solved for $\alpha_0$, the value of $\beta_0$ is determined by
\begin{equation}
\beta_0 = \left(\frac{c\Gamma(\gamma+2) 
\mathcal{L}_{\gamma+2}(e^{-\alpha_0})}
{E}\right)^{\frac{1}{\gamma+2}}=
\left(\frac{c\Gamma(\gamma+1) 
\mathcal{L}_{\gamma+1}(e^{-\alpha_0})}
{N}\right)^{\frac{1}{\gamma+1}}.
\label{betabar}
\end{equation}

Since all ingredients needed to build $\rho_{\rm mb}^{\so}$  are given by the above expressions, two comments are in order. First, we realize that the conjunction of the approximated formulas~(\ref{rhof}) and (\ref{smapp}) does not manifest an important behavior when $\alpha\to0$. If we just replace Eq.~(\ref{sbar}) on Eq.~(\ref{rho2x}) one verifies that the singular behavior evident on integral~(\ref{mainI}) in the limit $\alpha\to 0$ simply disappears. This problem clearly comes from the fact that the semiclassical approximation $\bar\rho_{\rm sp}$ leads to a non-accurate thermodynamical function $s$ in such a limit. Second, an inspection in the right hand side of Eq.~(\ref{alpha0}) shows that $\alpha_0$ is, in principle, completely determined if one knows $d$, that is, the mean-field potential, the number of particles $N$, and the total energy $E$. Extrapolating the validity limit of $\rho_{\rm mb}^{\so}$, we may notice that $d$ can assume values from 0 to $+\infty$. On the other hand, $\mathcal F_\gamma(\alpha_0)$ assumes values from $0$ to $\infty$, only when $-1<\gamma\le 0$. For $\gamma > 0$, however, we have
\begin{equation}
0<\mathcal F_\gamma(\alpha_0)< 
\frac{\zeta(\gamma+1)}
{\zeta(\gamma+2)^{\frac{\gamma+1}{\gamma+2}}}\equiv d^{\rm max}_{\gamma},
\label{alphacond}
\end{equation}
where the lower bound refers to $\alpha_0\to+\infty$, while the upper bound to $\alpha_0\to0$. Then, it is clear that, when $\gamma > 0$, a (real) saddle point ($\beta_0,\alpha_0$) will only exist if $d<d^{\rm max}_{\gamma}$. Out of this range, the approximation in the present form cannot be applied.

Finally, evaluation of $\rho_{\rm mb}^{\so}$ can be formally concluded, provided that Eq.~(\ref{alpha0}) has a solution. We simply write
\begin{equation}
\beta_0(\alpha_0) =  (\gamma+1)
\left[\mathcal L_{\gamma+2}(e^{-\alpha_0})\right]^{\frac{1}{\gamma+2}}
\left(\frac{N}{d\,E}\right).
\label{betabarQDL}
\end{equation}
Apart from this explicit dependence on $\alpha_0$, Eq.~(\ref{rhof}) can be finally written  as
\begin{equation}
\bar \rho^{\so}_{\mathrm{mb}}(E,N;\alpha_0) =
\frac{1}{\mathcal G}
\frac{\sqrt{\frac12\mathcal L_{\gamma+2}(e^{-\alpha_0}) }}
{2\pi E}
\exp{\left\{
(\gamma+2)\left[\mathcal L_{\gamma+2}(e^{-\alpha_0})
\right]^{\frac{1}{\gamma+2}}
\frac{N}{d}
+ \alpha_0N\right\}} ,
\label{rhofexp}
\end{equation}
where
\begin{equation}
\mathcal G \equiv
\frac{1}{\sqrt2}
\left|
\frac{\gamma+2}{\gamma+1}
\mathcal L_{\gamma}(e^{-\alpha_0})-
d^2
\left[\mathcal L_{\gamma+2}(e^{-\alpha_0})
\right]^{\frac{\gamma}{\gamma+2}}
\right|^{1/2}.
\end{equation}
Equation~(\ref{rhofexp}) is the final expression when formulas~(\ref{rhof}) and (\ref{smapp}) are considered. As discussed above, for $\gamma>0$, it can only be applied if $d<d^{\rm max}_{\gamma}$. Otherwise, no real saddle point $(\beta_0,\alpha_0)$ will exist in the integrand of Eq.~(\ref{rho2x}), with $s$ replaced by $\bar s$.

\subsection{Uniform approximation $\bar \rho^{\rm un}_{\mathrm{mb}}$}

Using the same assumption of the last section, namely, $ \bar\rho_{\mathrm{sp}}(\epsilon)=c\epsilon^\gamma$, we rewrite $\omega_e(\beta,\alpha)$ as
\begin{equation}
\begin{array}{lll}
\omega_e(\beta,\alpha) \approx \bar\omega_e(\beta,\alpha) 
&\equiv& \displaystyle
\left.\left(\frac{\bar n_{\mathrm{sp}}(\epsilon)
\ln(1-e^{-(\alpha+\beta\epsilon)})}{\beta}\right)
\right|_{\epsilon_1}^{\infty}
- \int_{\epsilon_1}^{\infty} 
\frac{\bar n_{\mathrm{sp}}(\epsilon)~d\epsilon}
{e^{\alpha+\beta\epsilon}-1} ,
\end{array}
\end{equation}
where $\epsilon_1=[(\gamma+1)/c]^{\frac{1}{\gamma+1}}$, once we have $\bar n_{\mathrm{sp}}(\epsilon_1)=2$. Then, the main ingredient $s_e$ to calculate $\bar \rho^{\rm un}_{\mathrm{mb}}$ can be written as
\begin{equation}
\begin{array}{lll}
\bar s_e(\beta,\alpha)&\equiv&\displaystyle
\beta E + \alpha N 
+\frac{c }
{\beta^{\gamma+1}}
\sum_{l=1}^{\infty}
\frac{e^{-l\alpha}
\overline\Gamma(\gamma+1;l\beta\epsilon_1)}{l^{\gamma+2}},
\end{array}
\label{se}
\end{equation}
where $\overline\Gamma$ is the incomplete Gamma function
\begin{equation}
\overline\Gamma(\nu+1;z) = \int_{z}^\infty x^\nu e^{-x}dx.
\end{equation}
Another important term can be straightforwardly derived,
\begin{equation}
\begin{array}{lll}
\displaystyle  \bar s_e^{(\beta\beta)} (\beta,\alpha)\equiv
\frac{\partial^2\bar s_e}{\partial\beta^2}&=&
\displaystyle
\frac{c\,\epsilon_1^{\gamma+2}}{\beta}
\mathcal L_0(e^{-(\alpha+\beta\epsilon_1)})+\frac{c(\gamma+2)}
{\beta^{\gamma+3}}
\sum_{l=1}^{\infty}\frac{e^{-l\alpha}
\overline\Gamma(\gamma+2;l\beta\epsilon_1)}{l^{\gamma+2}}.
\end{array}
\end{equation}
From Eq.~(\ref{se}), we can also write
\begin{equation}
\begin{array}{lll}
\bar{\mathcal{E}}_e(\beta,\alpha)&\equiv&\displaystyle
\int_{\epsilon_1}^{\infty}
\frac{c~\epsilon^{\gamma+1}~d\epsilon}{e^{\alpha+\beta\epsilon}-1}
=
\frac{c}{\beta^{\gamma+2}}
\sum_{l=1}^{\infty}\frac{e^{-l\alpha}
\overline\Gamma(\gamma+2;l\beta\epsilon_1)}{l^{\gamma+2}},\\
\bar{\mathcal{N}}_e(\beta,\alpha)&\equiv&\displaystyle
\int_{\epsilon_1}^{\infty}
\frac{c~\epsilon^{\gamma}~d\epsilon}{e^{\alpha+\beta\epsilon}-1}
= 
\frac{c}{\beta^{\gamma+1}}
\sum_{l=1}^{\infty}\frac{e^{-l\alpha}
\overline\Gamma(\gamma+1;l\beta\epsilon_1)}{l^{\gamma+1}}.
\end{array}\label{EeNe}
\end{equation}
From them, we can find the prescription to calculate $\tilde{\beta}_\alpha$, and consequently $\tilde{\beta}_0^*$, as well as the saddle point $(\tilde{\beta}_0,\tilde{\alpha}_0)$: 
\begin{equation}
\bar{\mathcal{E}}_e(\tilde\beta_\alpha,\alpha)=E, \quad
\bar{\mathcal{E}}_e(\tilde\beta_0^*,0)=E,  \quad
\bar{\mathcal{E}}_e(\tilde\beta_0,\tilde\alpha_0)=E, 
\quad\mathrm{and}\quad
\bar{\mathcal{N}}_e(\tilde\beta_0,\tilde\alpha_0)=N.
\label{saddletilde} 
\end{equation}
Differentiating the first of these equations and using the fact that $\delta E=0$, we find
\begin{equation}
\begin{array}{lll}
f(\alpha)&\equiv&
\displaystyle
\frac{\partial\tilde\beta_\alpha}{\partial\alpha}=
-\frac{c\,\tilde\beta_\alpha^{-(\gamma+1)}}{(\gamma+2)
\bar{\mathcal{E}}_e(\tilde\beta_\alpha,\alpha)+
c \epsilon_1^{\gamma+2} \mathcal L_0 (
e^{-(\alpha+\tilde\beta_\alpha\epsilon_1)})}
\sum_{l=1}^{\infty}\frac{e^{-l\alpha}
\overline\Gamma(\gamma+2;l\tilde\beta_\alpha\epsilon_1)}{l^{\gamma+1}},
\end{array}
\label{dbda}
\end{equation}
which is fundamental to evaluate the correction $\mathcal C$ present in Eq.~(\ref{unset}). To evaluate $\mathcal C$, it is also convenient to have an expression for $\sigma(\alpha)$. According to Eq.~(\ref{def1}) and using the first of Eqs.~(\ref{saddletilde}), we may approach it by
\begin{equation}
\begin{array}{lll}
\bar\sigma(\alpha)&\equiv&\displaystyle
\left(\frac{\gamma+2}{\gamma+1}\right)\tilde\beta_\alpha E + \alpha N 
-\mathcal L_1 (e^{-(\alpha+\tilde\beta_\alpha\epsilon_1)})-
\frac12 \ln \left[ \bar s_e^{(\beta\beta)}(\tilde\beta_\alpha,\alpha)
\right].
\end{array}
\end{equation}

\section{One-dimensional Harmonic Oscillator}
\label{OHS}

Now we are in position to apply the approximated formulas to a concrete test. The system of bosons confined to a one-dimensional harmonic oscillator (1D-HO) is ideal to play this role. Indeed, as this is the case where the energy levels are equidistant, the calculation of $\rho_{\mathrm{mb}}$ becomes identical to the problem of integer partition, which belongs to the field of number theory and has some analytical studies reported in the literature. This purely mathematical issue consists in finding the number of ways into which an integer $E$ (to make direct reference with our problem) can be expressed as a sum of, at most, $N$ other integer numbers. Our reference result in this context will be the formula due to Erdos and Lehner~\cite{erdos}, which is also valid only in the asymptotic limit of large values of $E$ and $N$. Our approach is equivalent to this when $c=1$ and $\gamma=0$, which characterize the 1D-HO spectrum. We will also have to assume large $E$ and $N$, but also $E\gg N$.

\subsection{Second order approximation $\bar{\rho}_{\mathrm{mb}}^{\so}$ -- 1D-HO}

Returning to Eq.~(\ref{alpha0}), we will look for values of $\alpha_0$ with the restriction $d=N^2/E\to\infty$. Equation~(\ref{alpha0}) becomes
\begin{equation}
\frac{\mathcal{L}_{1}(e^{-\alpha_0})}
{\sqrt{\mathcal{L}_{2}(e^{-\alpha_0})}}
\equiv
-\frac{\ln(1-e^{-\alpha_0})}
{\sqrt{\mathcal{L}_{2}(e^{-\alpha_0})}}
= \frac{N}{\sqrt{E}},
\label{saddle2}
\end{equation}
implying that $\alpha_0\approx e^{-N\sqrt{\zeta(2)/E}}$, if we simply use $\mathcal{L}_{2}(e^{-\alpha_0})\approx\zeta(2)=\pi^2/6$. Equation~(\ref{saddle2}) allows us to write $d$ as function of $\alpha_0$. If one replaces such a result in Eq.~(\ref{rhofexp}), one may expand it as
\begin{equation}
\bar \rho^{\so}_{\mathrm{mb}}(E,N;\alpha_0\to0) =
\frac{1}{2\pi E}
\sqrt{\frac12\zeta(2)\,\alpha_0}
\left[1+\mathcal{O}(\alpha_0)\right]
\exp{\left\{
\sqrt{E}\left[
2\sqrt{\zeta(2)}+\mathcal{O}(\alpha_0)
\right]
\right\}} \qquad (\mathrm{1D-HO}),
\label{rhobar}
\end{equation}
where now we approach $\mathcal{L}_{2}(e^{-\alpha_0})=\zeta(2)+\alpha_0\ln\alpha_0+\mathcal{O}(\alpha_0)$ and $\mathcal{L}_{0}(e^{-\alpha_0})=\frac{1}{\alpha_0}\left[1+\mathcal{O}(\alpha_0)\right]$. Finally, keeping only the leading correction in $\alpha_0\approx e^{-N\sqrt{\zeta(2)/E}}$, the many-body density of states becomes
\begin{equation}
\bar \rho^{\so}_{\mathrm{mb}}(E,N;\alpha_0\to0) =
\bar{\mathcal R}
\left\{
\frac{\sqrt{\frac12\zeta(2)}}
{2\pi E}
e^{2\sqrt{\zeta(2)E}} \right\}
\qquad (\mathrm{1D-HO}),
\label{rhofexpOHSQDL}
\end{equation}
with $\bar{\mathcal R}=e^{-\frac{N}{2} \sqrt{\frac{\zeta(2)}{E}}}$. The result of Erdos and Lehner~\cite{erdos}, in the extremal limit $N\to\infty$, which is also known as Hardy and Ramanujan formula~\cite{hardy}, is the expression seen {\em inside the brackets} of Eq.~(\ref{rhofexpOHSQDL}). Actually, the dependence on $N$ was included just in $\bar{\mathcal R}$, which clearly does not tend to 1 when $N\to\infty$. It shows that approximation~(\ref{rhofexp}) is ill-defined in the case $\alpha_0\to0$, or, equivalently, $N^2\gg E$. At last, for completeness, we note that, to be consistent with the work of Erdos and Lehner, the expression for $\bar{\mathcal R}$ in Eq.~(\ref{rhofexpOHSQDL}) should be 
\begin{equation}
\mathcal{R}_{\mathrm{EL}} = \exp\left\{ 
-\sqrt{\frac{E}{\zeta(2)}}~e^{-N\sqrt{\zeta(2)/E}}
\right\}.
\label{EL}
\end{equation}

\subsection{Uniform approximation $\bar{\rho}_{\mathrm{mb}}^{\mathrm un}$ -- 1D-OH}

For the case of the harmonic oscillator, Eqs.~(\ref{se}) and (\ref{EeNe}) become
\begin{equation}
\begin{array}{lll}
\bar s_e(\beta,\alpha)&=&\displaystyle
\beta E + \alpha N +\frac{1}{\beta}\mathcal L_2(e^{-(\alpha+\beta)}),\\
\bar{\mathcal{E}}_e(\beta,\alpha)&=&\displaystyle
\frac{1}{\beta^{2}}
\left[
\mathcal L_2(e^{-(\alpha+\beta)})+
\beta\mathcal L_1(e^{-(\alpha+\beta)})
\right],\\
\bar{\mathcal{N}}_e(\beta,\alpha)&=&\displaystyle
\frac{1}{\beta}\mathcal L_1(e^{-(\alpha+\beta)}).
\end{array}
\end{equation}
Then, $\tilde\beta_0^*$ should satisfy
\begin{equation}
\frac{1}{\left.\tilde\beta_0^*\right.^2}
\left[
\mathcal L_2(e^{-\tilde\beta_0^*})-
\tilde\beta_0^* \ln (1-e^{-\tilde\beta_0^*})
\right]=E.\label{b0s}
\end{equation}
A simple analysis of this equation shows that the quantity inside the brackets assumes values from $\zeta(2)$ to 0, when $\tilde\beta_0^*$ is varied from 0 to $+\infty$. Therefore, as $E$ is large, we conclude that the equality is satisfied for $\tilde\beta_0^*\ll1$, so that we can perform the straightforward expansions
\begin{equation}
\begin{array}{lll}
\bar s_e(\tilde\beta_0^*,0)&=&\displaystyle
\tilde\beta_0^* E + \frac{1}{\tilde\beta_0^*}\mathcal L_2(e^{-\tilde\beta_0^*})=
\frac{\zeta(2)}{\tilde\beta_0^*} + \ln\tilde\beta_0^* -1 +\left(\frac14+E \right)
\tilde\beta_0^*+ \mathcal{O}( \tilde{\beta}_0^{*^2}),\\
\bar s_e^{(\beta\beta)}(\tilde\beta_0^*,0)&=&\displaystyle
\frac{1}{\tilde\beta_0^*}
\mathcal L_0(e^{-\tilde\beta_0^*})+
\frac{2}{\tilde{\beta}_0^{*^3} } \left[
\mathcal L_2(e^{-\tilde\beta_0^*})
+\tilde\beta_0^* \mathcal L_1(e^{-\tilde\beta_0^*})\right]= 
2\frac{{\zeta(2)}}{{\tilde{\beta}^{*^3}_0}}\left[ 1+ \mathcal{O}(\tilde\beta_0^*)\right],\\
\end{array}\label{ssee}
\end{equation}
and also
\begin{equation}
\sigma_0^{(2)}=\frac{1}{{\beta_0^*}^2}
\left[ 1+\mathcal{O}(\beta_0^*)\right],\qquad
\sigma_0^{(3)}=-\frac{1}{{\beta_0^*}^3}
\left[ 1+\mathcal{O}(\beta_0^*)\right],
\qquad\mathrm{and}\qquad
\sigma_0^{(4)}=\frac{2}{{\beta_0^*}^4}
\left[ 1+\mathcal{O}(\beta_0^*)\right],
\end{equation}
so that, concerning $\mathcal C$, we have
\begin{equation}
c_0=\left( 1+\frac{1}{3\sqrt{2\pi}}\right)\left[ 1+\mathcal{O}(\tilde\beta_0^*)\right]
\quad\mathrm{and}\quad
c_1=\frac{25}{24}\sqrt{\frac{2}{\pi}}\frac{1}{\tilde\beta_0^*}\left[ 1+\mathcal{O}(\tilde\beta_0^*)\right].
\label{c0c1}
\end{equation}
To derive Eqs.~(\ref{ssee})-(\ref{c0c1}), we used 
\begin{equation}
\begin{array}{rll}
\bar s_e^{(\beta\beta)}(\beta,\alpha)&=&
\displaystyle
\frac{1}{\beta}
\mathcal L_0(e^{-(\alpha+\beta)})+
\frac{2}{\beta^{3}}\left[
\mathcal L_2(e^{-(\alpha+\beta)})
+\beta \mathcal L_1(e^{-(\alpha+\beta)})\right],\\
f(\alpha)&=&
\displaystyle
-\frac{1}{\tilde\beta_\alpha}
\frac{\displaystyle
\mathcal L_1 (e^{-(\alpha+\tilde\beta_\alpha)})
+\tilde\beta_\alpha \mathcal L_0 (e^{-(\alpha+\tilde\beta_\alpha)})}
{2\bar{\mathcal E}_e(\tilde\beta_\alpha,\alpha)+ 
\mathcal L_0 (e^{-(\alpha+\tilde\beta_\alpha)})},\\
\bar\sigma(\alpha)&=&\displaystyle
2\tilde\beta_\alpha E + \alpha N 
-\mathcal L_1 (e^{-(\alpha+\tilde\beta_\alpha)})-
\frac12 \ln \left[ \bar s_e^{(\beta\beta)}(\tilde\beta_\alpha,\alpha)
\right].
\end{array}
\end{equation}

At this point, we can already write
\begin{equation}
\bar \rho^{\mathrm{un}}_{\mathrm{mb}}(E,N;\tilde\alpha_0\to0) =
\frac{\mathcal C}{2\sqrt{2\pi}}
\left\{2\frac{{\zeta(2)}}{{\tilde{\beta}^{*^3}_0}}
\left[ 1+ \mathcal{O}(\tilde\beta_0^*)\right] \right\}^{-1/2}\tilde\beta_0^*
e^{\frac{\zeta(2)}{\tilde\beta_0^*} -1 +\left(\frac14+E \right)
\tilde\beta_0^*+ \mathcal{O}( \tilde{\beta}_0^{*^2})}
 \qquad (\mathrm{1D-HO}),
\label{rhobarun00}
\end{equation}
with 
\begin{equation}
{\mathcal C}=\left( 1+\frac{1}{3\sqrt{2\pi}}\right)\left[ 1+\mathcal{O}(\tilde\beta_0^*)\right]
\exp\left\{ 
-\left(\frac{25}{24}\sqrt{\frac{2}{\pi}}\frac{3\sqrt{2\pi}}{1+3\sqrt{2\pi}}\right)
\frac{\tilde\alpha_0}{\tilde\beta_0^*}
\left[ 1+\mathcal{O}(\tilde\beta_0^*)\right]
\right\}.
\end{equation}
To conclude, we just need to solve $\tilde\alpha_0$ and $\tilde\beta_0^*$. Expanding Eq.~(\ref{b0s}) around $\tilde\beta_0^*=0$, we find
\begin{equation}
\frac{\zeta(2)}{{\beta_0^*} ^2}-\frac{1}{{{\tilde\beta}_0^*}}+\frac14 +\mathcal{O}(\tilde\beta_0^*)=E
\quad\Longrightarrow \quad\tilde\beta_0^*= \sqrt{\frac{\zeta(2)}{E}} 
-\frac{1}{2E} +\mathcal{O}(E^{-3/2}),
\end{equation}
so that
\begin{equation}
\bar \rho^{\mathrm{un}}_{\mathrm{mb}}(E,N;\tilde\alpha_0\to0) =
\frac{\mathcal C}{2 e \sqrt{2\pi}}\sqrt{\frac{\zeta(2)^{3/2}}{2E^{5/2}}}
e^{2\sqrt{E \zeta(2)} }
\left[ 1+ \mathcal{O}(E^{-1/2})\right] 
 \qquad (\mathrm{1D-HO}).
\label{rhobarun0}
\end{equation}
The saddle point satisfies
\begin{equation}
\frac{1}{\tilde\beta_0^{2}}
\left[
\mathcal L_2(e^{-(\tilde\alpha_0+\tilde\beta_0)})-
\tilde\beta_0 \ln (1-e^{-(\tilde\alpha_0+\tilde\beta_0)})
\right]=E
\quad\mathrm{and}\qquad
-\frac{\ln (1-e^{-(\tilde\alpha_0+\tilde\beta_0)})}{\beta_0}=N.
\end{equation}
The last equation simply implies that $\tilde\alpha_0 = -\tilde\beta_0 -\ln[1-e^{-\tilde\beta_0 N}]$, which, replaced on the first equation, results in
\begin{equation}
\frac{\mathcal L_2(1-e^{-\tilde\beta_0 N})}{\tilde\beta_0^{2}}=E-N
\quad\Longrightarrow \quad \tilde\beta_0 \approx \sqrt{\frac{\zeta(2)}{E}}\left[ 1-\frac{N}{E}\right]^{-1/2},
\end{equation}
where we approach $\mathcal L_2(1-e^{-\tilde\beta_0 N})\approx\zeta(2)$,  that is consistent with the assumptions $N^2\gg E$ and $E\gg N$.  By inserting the last result in Eq.~(\ref{rhobarun0}) and disregarding the term $\frac{N}{E}$ above, we find
\begin{equation}
\bar \rho^{\mathrm{un}}_{\mathrm{mb}}(E,N;\tilde\alpha_0\to0) =
\bar{\mathcal R}_{\rm un}
\left\{
\frac{\sqrt{\frac12\zeta(2)}}
{2\pi E}
e^{2\sqrt{\zeta(2)E}} \right\}
 \qquad (\mathrm{1D-HO}),
\label{rhobarun}
\end{equation}
where
\begin{equation}
\mathcal{R}_{\mathrm{un}} = \frac{f_1}{E^{1/4}}\exp\left\{ 
-f_2\sqrt{\frac{E}{\zeta(2)}}~e^{- N\sqrt{\zeta(2)/E}}
\right\}.
\label{ELun}
\end{equation}
The quantities $f_1$ and $f_2$ are quantities containing only numerical terms, amounting to $f_1\approx1.23$ and $f_2\approx0.73$. Except for these numerical factors, that would need to be one to agree with Erdos and Lehner formula, two other mismatches have to be commented. First, to derive Eq.~(\ref{rhobarun}), we had to consider $E> N$, otherwise no real saddle point $\tilde\beta_0$ would exist. In addition, the condition $E\gg N$ was needed to reproduce result~(\ref{EL}). Second, the term $E^{-1/4}$ could not appear in the formula. However, as we will discuss in the following, such a problem can be solved if we improve the way in which information about the single-particle spectrum is furnished, which indicates that the cause of this problem lies on the smooth approximation and not on $\rho^{\mathrm{un}}_{\mathrm{mb}}(E,N;\tilde\alpha_0\to0)$.

\section{Final Remarks}
\label{fr}

In this paper, we consider a chain of approximating assumptions to develop an expression for $\rho_{\rm mb}(E,N)$. The first of them is to consider that the bosons are confined to a mean-field self-consistent potential, allowing us to write many-body spectrum in terms of the single-particle one. Second, we assume the thermodynamical limit, regime in which the integral representation of $\rho_{\rm mb}$ is proper to be approached by asymptotic expansion methods. For the 1D-HO, for instance, it is shown that the well-known second order saddle point method, which gives origin to $\rho_{\rm mb}^{\so}$, is not accurate when $N^2/E\to0$, condition reasonably true in the thermodynamical limit, because the saddle point approaches to a singular point of the integrand. We apply, therefore, another method to perform the integral and get a uniform formula $\rho_{\rm mb}^{\rm un}$ for the many-body density of states. Obviously, such a formula is demonstrated to be reduced to $\rho_{\rm mb}^{\so}$ when the saddle point is far from the mentioned singularity. By restricting only to the cases where the effects of the singularity are present, what can be considered our third assumption, we finally arrive to an expression for $\rho_{\rm mb}^{\rm un}(E,N;\tilde\alpha_0\to0)$, Eq.~(\ref{unset}).

With the uniform formula in hand, another approximation should be done in order to submit it to tests using the 1D-HO. This potential is specially useful for this issue, because of its connection with the problem of integer partition, which has solid analytical results for our reference. This fourth approximation inserted in the theory, in fact, refers to the way in which information about the single-particle spectrum is given. The natural choice is to use the Thomas-Fermi method, where $\rho_{\rm sp}$ usually becomes a power series. When it is done, one shows that the uniform formula $\rho_{\rm mb}^{\rm un}$ really improves $\rho_{\rm mb}^{\so}$ for the 1D-OH case. However, the dependence on $E$ of the pre-exponential factor does not match our reference result. In the following, we argue that the origin of this problem is not on $\rho_{\rm mb}^{\rm un}(E,N;\tilde\alpha_0\to0)$, but in the Thomas-Fermi approximation. Indeed, if one writes 
\begin{equation}
\begin{array}{lll}
s_e(\beta,\alpha) &=& \displaystyle
- \sum_{k=1}^{\infty} 
\ln(1- e^{-(\alpha+\beta k)}) + \beta E + \alpha N
\qquad (\mathrm{1D-HO}), 
\end{array}
\label{omegaex}
\end{equation}
the value of $s_e(\tilde\beta_0^*,0)$ can be calculated without the Thomas-Fermi approximation. Using the Dedekind eta function and its properties, we have~\cite{dedekind,bhaduri2004} 
\begin{equation}
\sum_{k=1}^{\infty}\ln\left[1-e^{-\tilde\beta_0^* k}\right] 
=
-\frac{\zeta(2)}{\tilde\beta_0^*} +
\frac12\ln\left[\frac{2\pi}{\tilde\beta_0^*}\right]
+\mathcal{O}(\tilde\beta^*_0).
\label{emseries}
\end{equation}
Using these last two equations just to write $s_e(\tilde\beta_0^*,0)$, the expression for $\bar \rho^{\mathrm{un}}_{\mathrm{mb}}(E,N;\tilde\alpha_0\to0)$ assumes the same form as Eq.~(\ref{rhobarun}), except for $\mathcal R_{\rm un}$, which becomes
\begin{equation}
\mathcal{R}'_{\mathrm{un}} = f_3\exp\left\{ 
-f_2\sqrt{\frac{E}{\zeta(2)}}~e^{- N\sqrt{\zeta(2)/E}}
\right\}.
\end{equation}
where $f_3\approx 1.04$. It then demonstrates that the uniform formula considerably improves $\rho^{\so}_{\mathrm{mb}}$, given its agreement with the Erdos and Lehner formula. At last, it should be commented that Tran {\it et al}~\cite{bhaduri2004} also derived approximations for $\rho_{\mathrm{mb}}$ using a different (but similar) route. In their work, the Thomas-Fermi method is not used to provide information about the single-particle spectrum. Their choice is dealing with spectra given by a power-law $\epsilon_i = i^{r}$, performing the sum present in the definition of $s$ using methods that here were applied to Eq.~(\ref{emseries}). Although it avoids the discrepancy generated by the Thomas-Fermi method, its disadvantage resides on the fact that finite $N$ corrections can be easily found only for $r=1$ (1D-HO case).

\section*{Acknowledgments}

This work started when the author spent a few months in the group of Patricio Leboeuf at the Laboratoire de Physique Th\'eorique et Mod\`eles Statistiques (LPTMS), Universit\'e Paris 11. The author is grateful for the hospitality in this period and the scientific discussions with the group. Financial support provided by CAPES, CNPq, and the National Institute for Science and Technology of Quantum Information (INCT-IQ) is also acknowledged.

\appendix
\section{Jacobian at the critical points}
\label{app1}

Our task here is finding expressions for $G(0)$ and $G(t_0)$, where $G(t)$ is given by Eq.~(\ref{jacobG}). The function $G(t)$ involves the term $t/(1-e^{-\alpha(t)})$, which becomes undetermined by directly replacing $t=0$. Thus, we apply the L'Hospital's rule on it obtaining
\begin{equation}
\lim_{t\to0}
\left(\frac{t}{1-e^{-\alpha(t)}}\right) = 
\lim_{t\to0}
\left(\frac{1}{d\alpha/dt}\right) .
\end{equation}
Therefore,
\begin{equation}
G(0)=
\lim_{t\to 0}\left\{
\frac{t}{1-e^{-\alpha(t)}} 
\frac{d\alpha}{dt}\right\}= 1.
\label{g0}
\end{equation}
In order to find an expression for $G(t_0)$, we differentiate Eq.~(\ref{map}), finding
\begin{equation}
\frac{d\sigma}{d\alpha}
\frac{d\alpha}{dt} = \frac{d\phi}{dt}
\quad\Longrightarrow \quad
\frac{d\alpha}{dt} = 
-\frac{\lambda(t+\gamma)}
{d\sigma/d\alpha},
\label{d2phi}
\end{equation}
Again, to find the value of the last equality when $t=t_0=-\gamma$ we need to apply the L'Hospital's rule,
\begin{equation}
\lim_{t\to t_0}
\left(\frac{d\alpha}{dt}\right) = 
\lim_{t\to t_0}\left(
-\frac{\lambda}
{\left(d^2\sigma/d\alpha^2\right)
\left(d\alpha/dt\right)}\right),
\end{equation}
implying that
\begin{equation}
\lim_{t\to t_0}
\left(\frac{d\alpha}{dt}\right)^2 = -
\frac{\lambda}
{\left.\left(d^2\sigma/d\alpha^2
\right)\right|_{\alpha=\tilde\alpha_0} }.
\end{equation}
Therefore, 
\begin{equation}
G(t_0) = \pm
\frac{i\,t_0}{1-e^{-\tilde\alpha_0}}
\left[ 
\frac{\lambda}
{\left.\left(d^2\sigma/d\alpha^2
\right)\right|_{\alpha=\tilde\alpha_0} }\right]^{1/2},
\label{gt0}
\end{equation}
where the choice of the sign can be done by imposing that 
\begin{equation}
\lim_{t_0\to0}G(t_0)=G(0)=1.
\end{equation}
To accomplish this task, we expand Eq.~(\ref{Dsigma}) around $\tilde\alpha_0=0$,
\begin{equation}
\begin{array}{lll}
\displaystyle t_0\sqrt\lambda &=& \displaystyle
-\gamma\sqrt\lambda =  - 
\sqrt{2[\sigma(\tilde\alpha_0)-\sigma(0)]}\\
&=&\displaystyle 
-i\left\{
\sigma^{(2)}_0\tilde\alpha_0^2
\left[
1+\frac{2}{3} \frac{\sigma^{(3)}_0}{\sigma^{(2)}_0}\tilde\alpha_0
+\frac{1}{4} \frac{\sigma^{(4)}_0}{\sigma^{(2)}_0}\tilde\alpha_0^2
+\mathcal{O}(\tilde\alpha_0^3)
\right]
\right\}^{1/2},
\end{array}
\label{exp1}
\end{equation}
where we use the notation $\sigma^{(k)}_0\equiv (d^k\sigma/d\alpha^k)|_{\alpha=0}$. To find the last expression, we point out that we eliminate the term $\sigma^{(1)}_0$ from the expansion because it vanishes when $\tilde\alpha_0\to0$. To do so, we use
\begin{equation}
\left.\left(\frac{d\sigma}{d\alpha}
\right)\right|_{\alpha=\tilde\alpha_0} = \sigma^{(1)}_0+
\sigma^{(2)}_0\tilde\alpha_0+\frac12\sigma^{(3)}_0\tilde\alpha_0^2
+\mathcal{O}(\tilde\alpha_0^3) = 0.
\end{equation}
Then, using Eq.~(\ref{exp1}) we finally get
\begin{equation}
\lim_{t_0\to0} G(t_0)=\lim_{\tilde\alpha_0\to0}\left\{
\pm 
\frac{i\,t_0\sqrt\lambda}{\tilde\alpha_0+\mathcal{O}(\tilde\alpha_0^2)}
\left[
\frac{1}
{\left.\left(d^2\sigma/d\alpha^2
\right)\right|_{\alpha=\tilde\alpha_0} }\right]^{1/2}
\right\}=\pm1,
\end{equation}
implying that the plus sign should be chosen in Eq.~(\ref{gt0}).

\section{Derivatives of $\sigma(\alpha)$ in terms of $s_e(\beta,\alpha)$}
\label{app3}

In this appendix, we will write derivatives of $\sigma(\alpha)$ in terms of derivatives of $s_e(\beta,\alpha)$, where $\beta=\beta(\alpha)$. For a generic point $(\beta,\alpha)$, up to leading order on $\lambda$, we have 
\begin{equation}
\frac{d\sigma}{d\alpha} = 
\frac{\partial s_e}{\partial\beta}\frac{\partial\beta}{\partial\alpha} + 
\frac{\partial s_e}{\partial\alpha}
\end{equation}
and
\begin{equation}
\begin{array}{lll}
\displaystyle
\frac{d^2\sigma}{d\alpha^2} &=& 
\displaystyle
\frac{\partial }{\partial\beta}\left[\frac{\partial s_e}
{\partial\beta}\frac{\partial\beta}{\partial\alpha}\right]
\frac{\partial\beta}{\partial\alpha} +
\frac{\partial }{\partial\alpha}\left[\frac{\partial s_e}
{\partial\beta}\frac{\partial\beta}{\partial\alpha}\right]
+
\frac{\partial^2s_e}{\partial\beta\partial\alpha}
\frac{\partial\beta}{\partial\alpha}+
\frac{\partial^2s_e}{\partial\alpha^2},
\end{array}
\label{b2}
\end{equation}
where the first two terms of the right hand side can be written as
\begin{equation}
\begin{array}{l}
\displaystyle
\frac{\partial^2s_e}{\partial\beta^2}
\left(\frac{\partial\beta}{\partial\alpha} \right)^2 +
\frac{\partial s_e}{\partial\beta}
\left(
\frac{\partial^2\beta}{\partial\beta\partial\alpha}
\frac{\partial\beta}{\partial\alpha}+
\frac{\partial^2\beta}{\partial\alpha^2}
\right)+
\frac{\partial^2s_e}{\partial\alpha\partial\beta}
\frac{\partial\beta}{\partial\alpha}.
\end{array}
\label{b3}
\end{equation}
Remembering that to find $\beta(\alpha)$ we impose $\partial s_e/\partial\beta=0$, we conclude that, under this constraint,
\begin{eqnarray}
\frac{d\sigma}{d\alpha} = \frac{\partial s_e}{\partial\alpha}
\end{eqnarray}
and also that Eq.~(\ref{b3}) is zero, namely,
\begin{equation}
\begin{array}{lll}
\displaystyle
\frac{\partial^2s_e}{\partial\beta^2}
\left(\frac{\partial\beta}{\partial\alpha} \right)^2 +
\frac{\partial^2s_e}{\partial\alpha\partial\beta}
\frac{\partial\beta}{\partial\alpha} = 0
& \Longrightarrow &\displaystyle
\frac{\partial\beta}{\partial\alpha} = -
\frac{\displaystyle
\frac{\partial^2s_e}{\partial\alpha\partial\beta}}
{\displaystyle
\frac{\partial^2s_e}{\partial\beta^2}}.
\end{array}
\label{b5}
\end{equation}
Therefore,
\begin{equation}
\begin{array}{lll}
\displaystyle
\frac{d^2\sigma}{d\alpha^2}&=&
\displaystyle
\frac
{\displaystyle
\frac{\partial^2s_e}{\partial\alpha\partial\beta}\frac{\partial^2s_e}{\partial\alpha\partial\beta}-
\frac{\partial^2s_e}{\partial\alpha^2}\frac{\partial^2s_e}{\partial \beta^2} }
{\displaystyle
\frac{\partial^2s_e}{\partial \beta^2}}.
\end{array}
\label{C6}
\end{equation}


\end{document}